\documentclass[preprint,12pt]{elsarticle}

\usepackage[left=2.5cm,right=2.5cm,top=2.5cm,bottom=3cm]{geometry}

\usepackage{amssymb}

\biboptions{sort&compress}

\newcounter{bla}

\usepackage{algorithm}
\usepackage{algpseudocode}

\usepackage{amsmath}
\usepackage{hyperref}
\newcommand{\code}[1]{\texttt{#1}}
\newcommand{\FFSol}{\texttt{Finite\-Field\-Solve}}
\newcommand{\LinSysSol}{\texttt{Linear\-System\-Solver}}
\newcommand{\FiniteFlow}{\texttt{Finite\-Flow}}
\newcommand{\ffdense}{\code{FFDenseSolve}}
\newcommand{\ffsparse}{\code{FFSparseSolve}}
\newcommand{\OL}{\texttt{Options\-List}}
\newcommand{\kira}{\code{Kira}}
\newcommand{\spasm}{\code{SpaSM}}
\newcommand{\spasmlink}{\code{spaSMLink}}
\newcommand{\firefly}{\code{FireFly}}
\newcommand{\fermat}{\code{Fermat}}
\newcommand{\Zp}{\mathbb{Z}_p}
\usepackage{tabularx}
\usepackage{caption}
\usepackage{subcaption}
\usepackage{orcidlink}

\usepackage{tikz-feynman}

\journal{Computer Physics Communications}

\begin{document}

\begin{frontmatter}



\title{\texttt{FiniteFieldSolve}: Exactly Solving Large Linear Systems in High-Energy Theory}


\author[a]{James Mangan\corref{author} \orcidlink{0000-0002-9713-7446}}

\cortext[author] {Corresponding author.\\\textit{E-mail address:} james.mangan@northwestern.edu}
\address[a]{Department of Physics and Astronomy, Northwestern University, Evanston, Illinois 60208, USA}

\begin{abstract}

Large linear systems play an important role in high-energy theory, appearing in amplitude bootstraps and during integral reduction.
This paper introduces \texttt{FiniteFieldSolve}, a general-purpose toolkit for exactly solving large linear systems over the rationals.
The solver interfaces directly with Mathematica, is straightforward to install, and seamlessly replaces Mathematica's native solvers.
In testing, \texttt{FiniteFieldSolve} is approximately two orders of magnitude faster than Mathematica and uses an order of magnitude less memory.
The package also compares favorably against other public solvers in \texttt{FiniteFieldSolve}'s intended use cases.
As the name of the package suggests, solutions are obtained via well-known finite field methods.
These methods suffer from introducing an inordinate number of modulo (or integer division) operations with respect to different primes.
By automatically recompiling itself for each prime, \texttt{FiniteFieldSolve} converts the division operations into much faster combinations of instructions, dramatically improving performance.
The technique of compiling the prime can be applied to any finite field solver, where the time savings will be solver dependent.
The operation of the package is illustrated through a detailed example of an amplitude bootstrap.

\end{abstract}


\end{frontmatter}



{\bf PROGRAM SUMMARY/NEW VERSION PROGRAM SUMMARY}

\begin{small}
\noindent
{\em Program Title:} \texttt{FiniteFieldSolve}                                          \\
{\em CPC Library link to program files:} (to be added by Technical Editor) \\
{\em Developer's repository link:} \url{https://github.com/jfmangan/FiniteFieldSolve} \\
{\em Code Ocean capsule:} (to be added by Technical Editor)\\
{\em Licensing provisions:} GPLv3  \\
{\em Programming language:} Mathematica, C++                                \\
{\em Nature of problem:} Exactly solving large linear systems over the rationals occurs in various settings in high-energy theory, for example when performing integral reduction or bootstrapping an amplitude.\\
{\em Solution method:} The linear system is solved by repeatedly row reducing over different finite fields (see Ref [1] and references therein).
Finite fields avoid the intermediary expression swell inherent to arbitrary precision rationals and bypass roundoff errors from floating point numbers.
A downside to using modular arithmetic is that it introduces a tremendous number of integer divisions, but this can be mitigated by compiling the divisions down to simpler instructions.
The solver is designed to handle arbitrarily dense systems such as those that appear in certain amplitudes bootstraps.\\
   \\

\end{small}


\section{Introduction}\label{sec:intro}

Exactly solving large linear systems is a key tool in the theorist's arsenal.
The most time consuming and computationally expensive steps in many quantum field theory calculations are integral reduction and integration, where the former can be reduced to a linear algebra problem \cite{Laporta:2000dsw}.
Large linear systems also appear prominently in amplitude bootstraps, including the double copy \cite{Bern:2010ue, Bern:2008qj}, the soft bootstrap \cite{Cheung:2014dqa}, as well as others.
The double copy in particular has been crucial in understanding the ultraviolet behavior of supergravity \cite{Bern:2018jmv, Bern:2017ucb, Bern:2014sna, Bern:2013uka, Bern:2012gh, Bern:2012cd, Bern:2012uf, Bern:2011qn} and in making precision gravitational wave predictions \cite{Cheung:2018wkq, Bern:2019nnu, Bern:2019crd, Bern:2021dqo}.
There are a number of publicly distributed linear solvers \cite{eigen, flint, linbox, spasm, spasmlink, LinSysSol}, including some integrated into integral reduction software \cite{Smirnov:2019qkx, vonManteuffel:2012np, Klappert:2020nbg}.
However, interfacing with these solvers can be a non-trivial task and some of them are intended purely for very sparse systems.
This paper presents \FFSol{}, a general-purpose exact large linear system solver for arbitrary density equations over the rationals that can be called directly from Mathematica.
The solver presented in this paper is open source, a drop in replacement for Mathematica's \code{Solve} and \code{Reduce}, easy to install and use, and roughly two orders of magnitude faster than Mathematica and uses an order of magnitude less memory in applications tested (see Tables \ref{tbl:benchmark1}, \ref{tbl:benchmark2}, and \ref{tbl:benchmark3}).\footnote{The author is only aware of three other solvers that link directly into Mathematica and are intended for exact large linear systems over the rationals.
The first is \LinSysSol{} \cite{LinSysSol} based on \cite{Kauers:2008zz}, the second is \FiniteFlow{} \cite{finiteflow} based on \cite{Peraro:2019svx}, the and the third is \spasmlink{} \cite{spasmlink} which is based on \spasm{} \cite{spasm}.
Some of the frontend of \FFSol{} was adapted from \spasmlink{} but \FFSol{} is unrelated to \spasm{}.}
Although \FFSol{} is designed for arbitrarily dense systems, some of the methods presented here are applicable to other solvers, including the technique of compiling the prime as described below.
\FFSol{} also has the tremendous advantage that its memory footprint is very stable and predictable and the time to row reduce a matrix over a given finite field can be estimated very accurately.
This makes \FFSol{} highly predictable while providing users access to a wide range of physics problems that would otherwise be intractable, all with the convenience of Mathematica's high-level syntax.

As a general-purpose solver, \FFSol{} is equipped to handle fully dense systems.
While sparse solvers will outperform \FFSol{} for very sparse systems, the ability to manipulate high density equations allows the user to offload certain challenges to the computer.
Specifically, the user is no longer responsible for carefully generating sparse systems.
For example, imposing factorization on the ansatz for some scattering amplitude naturally involves rational functions of kinematics.
Combining fractions in order to obtain sparse equations can become prohibitively complicated and is difficult to parallelize.
Instead, the equations can be extracted by repeatedly sampling random configurations of the kinematics, \emph{i.e.}, by plugging in random integers for the generalized Mandelstams.
While this will generally produce dense equations, it is trivial to parallelize, and all of the complications of solving the system can be foisted onto \FFSol{}.
High density equations can also show up in bootstraps of scalar theories since there is no helicity structure that breaks the equations into sparse sub-sectors.

While \FFSol{} can natively solve dense systems, some strategies employed by the solver are not restricted to such systems.
Understanding these strategies requires a brief foray into finite field methods.
The standard approach to exactly solving linear systems is to repeatedly row reduce the coefficient matrix over different primes and then reconstruct the full rational solution at the end.
Modular arithmetic avoids the round off errors inherent to floats and is impervious to the intermediary expression swell that plagues arbitrary precision rationals.\footnote{Expression swell should eventually subside since the final answer is expected to be simple on physical grounds.}
However, there are two main drawbacks to finite fields.

First, modular arithmetic introduces a tremendous number of integer division (or modulo) operations, which is by far the slowest of the four basic arithmetic operations.
This issue can be dramatically improved by ``compiling'' the prime.
\FFSol{} uses the compiler's optimizer to convert modulo operations with respect to a constant prime into much faster combinations of simpler instructions.
This means that the row reduction algorithm must be \emph{recompiled} for each prime, but this up front cost is negligible for large systems.
Compiling the prime results in overall performance increasing by up to a factor of 4 in testing.
All solvers that rely on modular arithmetic can benefit from compiling the primes, but the performance gain will depend on how much time the solver spends on actual arithmetic as opposed to memory allocation or pivot selection.

The second issue with finite field methods is that the matrix must be solved multiple times over different primes in order to reconstruct the full rational solution.
This can lead to extraneous work if large portions of the answer require only a few primes to reconstruct but certain pieces of the answer require many primes to capture.
\FFSol{} attempts to minimize wasted effort by gleaning as much information as possible from the first row reduction.
For example, with high probability, it only takes one or a small number of primes to determine:  the linearly independent equations in a system, the rank of the system, which variables are (in)dependent, which variables are set to zero, if the system is inconsistent, etc.
This information can be reused over subsequent primes to reduce the size of the matrix that needs to be row reduced, which can dramatically improve performance as the time to row reduce an $n\times n$ matrix is naively $\mathcal{O}(n^3)$.
Wisely reusing information from the first few primes can be applied to all finite field solvers and is already in use in at least some \cite{Klappert:2020nbg, Peraro:2019svx}.

Section \ref{sec:FFMethods} gives a brief review of finite field methods in the context of exactly solving linear systems.
Technical details of \FFSol{} are presented in Section \ref{sec:technical} along with comments about which techniques in \FFSol{} may be applicable to other solvers.
Besides compiling the primes and reusing information from early primes, \FFSol{} improves performance by: using two forms of parallelization, statically allocating memory for faster access, and using 16-bit primes to improve speed and memory consumption.
Section \ref{sec:guide} explains installing \FFSol{}, which amounts to running an installer script, installing a compiler if necessary, and downloading an example file if desired.
Section \ref{sec:guide} also provides a guide to the most important, high-level functions in \FFSol{}.
Example calculations and benchmarks against other solvers are provided in Section \ref{sec:example}.
The section includes a detailed example of bootstrapping the 8pt Dirac-Born-Infeld (DBI) tree amplitude \cite{Cheung:2014dqa} as well as an overview of constructing the color-dual 4pt two-loop non-linear sigma model (NLSM) integrand \cite{Edison:2023ulf}.
The examples demonstrate a workflow for \FFSol{} where constraints are implemented one at a time and the linear system is periodically pruned of unnecessary information.
Section \ref{sec:conclusions} contains some concluding remarks.

\section{Finite field methods}
\label{sec:FFMethods}

Finite fields are a standard tool in exactly solving linear systems so only a brief review is presented here.
For background and related material see Refs \cite{Kauers:2008zz, vonManteuffel:2014ixa, Cohen_2000, Hardy_Wright_2008, Peraro:2016wsq, Peraro:2019svx, Magerya:2022hvj, Klappert:2019emp, Zippel, Demillo, Schwartz} and the references therein.
Finite fields are the method of choice because they are unaffected by intermediary expression swell and roundoff errors.
The tradeoff is that the solutions produced are only probabilistically correct.
At a very high level, these solvers work by:  first row reducing the linear system over different finite fields $\Zp$, then stitching together the results with the Chinese remainder theorem (CRT), and then finally reconstructing the rational solution using the extended Euclidean algorithm (EEA).
For convenience, the relevant number theoretic algorithms are summarized in \ref{sec:appendix}.

In more detail, \FFSol{} begins by converting an inhomogeneous system of equations $A\mathbf{x}=\mathbf{b}$ into a homogeneous one by appending $\mathbf{-b}$ to $A$ and solving
\begin{align}
(A, \mathbf{-b}) \begin{pmatrix} \mathbf{x}\\ x_0 \end{pmatrix} =0,
\end{align}
where $x_0$ is taken to be 1 at the end of the calculation.
The system is inconsistent if solving the system sets $x_0$ to a particular value.
In this way any problem can be converted to a homogeneous one of the form $A \mathbf{x}=0$.
The approach to solving this homogeneous system is summarized in Algorithm \ref{alg:RowReduce}.
In order to solve the system, the matrix $A$ is projected onto a finite field $\Zp$ where $p$ is prime.
Any denominators in $A$ can be projected onto $\Zp$ using Fermat's little theorem.
Of course, if $A$ contains any factors of $1/p$ then a different prime must be selected, but this is relatively rare in practice.
The main algorithm loops over primes $p_1, p_2...$ starting at $p_1$.
Using any desired row reduction algorithm, the system $A \mathbf{x}_{p_1} \equiv 0 \mod p_1$ is solved for $\mathbf{x}_{p_1}$.
(In Algorithm \ref{alg:RowReduce} and what follows, $\mathbf{x}_a$ denotes the solution to $A \mathbf{x}_a \equiv 0 \mod a$ where $a$ could be any integer.)
The algorithm then attempts to reconstruct the rational solution $\mathbf{x}$ to $A\mathbf{x}=0$ using the extended Euclidean algorithm (EEA).
A fraction $a/b$ will be successfully reconstructed when $|a|$ and $b$ are smaller than about $\sqrt{p_1}$ \cite{RR1, RR2, RR3}.
Since larger primes allow more fractions to be reconstructed, conventional wisdom is to use the largest primes that fit in the word size of the machine.
Counterintuitively, \FFSol{} reaps considerable performance benefits from choosing slightly smaller primes.
In any case, if reconstruction fails, then the algorithm proceeds to solve $A \mathbf{x}_{p_2} \equiv 0 \mod p_2$ over a second prime.
Using the Chinese remainder theorem (CRT), the information from $\mathbf{x}_{p_1}$ and $\mathbf{x}_{p_2}$ is combined to obtain $\mathbf{x}_{p_1 \cdot p_2}$, which solves $A \mathbf{x}_{p_1 \cdot p_2} \equiv 0 \mod (p_1 p_2)$ by construction.
Using $\mathbf{x}_{p_1 \cdot p_2}$ is enough to reconstruct rational solutions with numerators and denominators less than about $\sqrt{p_1 p_2}$.
The algorithm row reduces $A$ over more and more primes until $\mathbf{x}$ solves the original (rational) problem $A \mathbf{x}=0$.
If primes $p_1, p_2...p_i$ have been used, it is often sufficient to check that $\mathbf{x}$ is a solution over the next prime that would be used, \emph{i.e.}, $A \mathbf{x} \equiv 0 \mod p_{i+1}$.
With enough primes it is possible to reconstruct arbitrarily complicated rational solutions.\footnote{Although the EEA is sufficient for reconstructing rational solutions, more specialized algorithms exist that attempt to limit the number of required primes.  One example is the algorithm in Ref \cite{RR3} which has been implemented in \firefly{}.  The author would like to thank an anonymous referee for pointing this out.}
In practice, typically only a few primes are needed because the rational numbers in physics problems tend to be simpler than those that come from solving a random matrix.

\begin{algorithm}
\caption{Solve $A\mathbf{x}=0$}\label{alg:RowReduce}
\begin{algorithmic}
\Require Matrix $A$ and primes $p_1, p_2...$
\Ensure General solution $\mathbf{x}$ satisfying $A\mathbf{x}=0$
\State $i \gets 1$
\Repeat
\State Solve $A \mathbf{x}_{p_i} \equiv 0 \mod p_i$ for $\mathbf{x}_{p_i}$
\State Build $\mathbf{x}_{p_1\cdot p_2...p_i}$ from CRT on $\mathbf{x}_{p_1 \cdot p_2... p_{i-1}}$ and $\mathbf{x}_{p_i}$
\State Reconstruct rational $\mathbf{x}$ from EEA on $\mathbf{x}_{p_1\cdot p_2...p_i}$
\State $i \gets i+1$
\Until{$A\mathbf{x} = 0$}
\end{algorithmic}
\end{algorithm}

\section{Technical details}
\label{sec:technical}

\FFSol{} implements the solver described in the previous section through two distinct components:  a high-level frontend and a low-level backend.
The backend is responsible for performing the actual row reduction over a prime.
As this is the most computationally intensive part of the algorithm, the backend is written in C++ to be as performant as possible.
The frontend of \FFSol{} is a Mathematica wrapper that interfaces with the row reducer and is responsible for reconstructing the full rational solution.
The core structure of the front end is based on \spasmlink{} \cite{spasmlink} but it has been dramatically altered to include the features mentioned in this section.

The performance of the backend comes from leveraging modern CPU architecture and optimizing compilers.
The C++ backend employs a modified version of ordinary Gauss-Jordan elimination with unsigned 16-bit integers and is \emph{recompiled} each time it is used.
These unusual design features are centered around the following five performance considerations.
\begin{enumerate}
\item Gauss-Jordan elimination is a highly parallelizable algorithm since many row operations can be performed independently.
Multi-core parallelization comes with a small overhead but the testing in Section \ref{sec:example} demonstrates that the benefits outweigh the costs, sometimes by a generous margin.
\FFSol{} uses OpenMP for parallelization where more details are given in Section \ref{sec:guide}.
Unlike more specialized algorithms, Gauss-Jordan elimination is highly predictable, so it is easy to reliably bound the time it will take to solve a system, at least over a single prime.
\item Gauss-Jordan elimination also benefits from vectorization where multiple elements in a row are updated simultaneously using the SIMD instruction set extension of the processor.
For x86 processors, AVX2 is the relevant instruction set extension since it provides SIMD instructions for \emph{integer} registers.
 \item Representing the matrix over a finite field incurs a tremendous number of modulo (or division) operations during row reduction.
 Since division is both slow and common in this algorithm, it is a prime target for optimization.
 The frontend of \FFSol{} automatically recompiles the backend each time the row reduction routine is called, effectively making the prime a compile-time constant.
 The optimizer can then turn divisions with respect to the prime into much faster combinations of instructions, significantly improving overall performance \cite{GranlundMontgomery}.\footnote{If the divisor is known at compile time, gcc uses the methods of Ref \cite{GranlundMontgomery} to optimize the division.  Ref \cite{2019arXiv190201961L} proposed a faster method if just the remainder of the division is required, but gcc does not appear to make use of this approach.  If the divisor is not known at compile time, there are still methods for optimizing the divisions \cite{libdivide}.}
Mersenne and Mersenne-like primes compile down to simpler instructions, but this does not appear to result in improved real-world performance, likely due to cache misses and memory bandwidth limitations.
As will be shown in Section \ref{sec:example}, the single decision to compile the prime leads to overall performance increasing by up to a factor of 4.
\item By virtue of recompiling the backend for each row reduction, the matrix dimensions are also known at compile time, allowing the compiler to statically allocate memory for the matrix.
Statically allocated memory is typically much faster to manipulate than dynamically allocated memory.
Forcing the backend to dynamically allocate the matrix results in overall performance dropping by a factor of 2 to 3 in testing.
\item Unsigned 16-bit integers are used for representing elements of the finite field.
16-bit integers have the advantage that the product of two such integers fits in a machine word, avoiding complicated overflow logic, and 16 bits is enough to reconstruct many of the simple fractions like 1/2 that pervade physics.
Choosing small integers improves the memory footprint of \FFSol{} -- since multiple integers can be packed into a machine word -- and thereby improves performance by alleviating memory bandwidth bottlenecks.
16-bit integers also happen to be faster than their 32-bit counterparts.
Because 32-bit integers are twice as long, it typically requires half as many primes to reconstruct the same rational solution.
To make up for this, 32-bit integers would only need to be half as fast as 16-bit integers, but this is not the case in testing.
Likewise, 8-bit integers would need to be twice as fast as 16-bit integers, but again this is not borne out in testing.
16-bit integers thus achieve the best balance of speed and size.
\FFSol{} uses 16-bit integers by default but there are options to use 32-bit integers if desired.
\end{enumerate}

The five considerations above contribute substantially to the performance of the backend but there are several features of the Mathematica frontend that also increase performance.
Understanding these features requires a brief detour into fill-in and pivoting.
The backend uses the first available pivot, working its way from left to right and top to bottom in the matrix.
This has the advantage that the algorithm cannot stall in the process of searching for optimal pivots as can occur for more complicated pivoting approaches.
However, this naive pivoting algorithm may incur serious fill-in.
As a simple but effective measure to mitigate fill-in, the frontend sorts the rows of the matrix by their density so that the sparsest rows are at the top of the matrix.
Although fill-in will not negatively impact memory usage, as it would in a sparse algorithm, minimizing fill-in improves performance since entire rows can be skipped if they have a zero in the pivot column.
Said another way, the frontend attempts to avoid situations where a very dense row at the top of the matrix would contaminate subsequent sparse rows.
The frontend also improves performance by using information from the first few primes to shrink the matrix it must solve over subsequent primes.
Specifically, the frontend uses the first prime to find the linearly independent equations and only solves that subset of equations over the remaining primes.
Last, the row reduction over the first prime provides a good indication of which variables the equations set to zero, allowing the frontend to limit the number of columns used with future primes.

So far, this section has mostly focused on the techniques used to improve the speed of \FFSol{}.
The discussion would not be complete without an overview of the memory layout of the package as memory can become the limiting resource for large matrices.
Care was taken in \FFSol{} to reduce the memory footprint and produce stable, predictable memory usage, so that the package is predictable in both time and memory.
The majority of the memory is used for representing three matrices.
First, Mathematica stores the original input matrix as a \code{List} or \code{SparseArray} of arbitrary precision rationals.
The total memory used to represent this matrix varies based on its dimensions and density but is constant in time.
The second matrix is a copy of the first one after reduction modulo a prime $p$.
The C++ backend operates on this second matrix.
To conserve memory, the matrix is passed from the frontend to the backend one row at a time.
When \FFSol{} is used in its default 16-bit mode, this $m \times n$ matrix requires $2mn$ bytes of memory.
Since the solver tries to eliminate unnecessary rows and columns when operating on subsequent primes, the memory needed to store the backend matrix tends to decrease over time.
For anything except the densest input systems, the majority of the memory will be used by the 16-bit backend matrix.
The third and final matrix is used to represent the rational solution to the input system.\footnote{At various stages in the calculation the solution is actually made of multiple matrices that are combined using the Chinese remainder theorem.  However, this does not alter the overall picture of the memory usage.}
This third matrix is stored in Mathematica as a \code{SparseArray} of arbitrary precision rationals.
Since a generic solution looks schematically like a diagonal of ones with some number of non-zero columns, this matrix tends to be very sparse and consumes the least memory of any of the matrices.
However, the memory used to store this matrix will increase slowly over time as the rational numbers inside of it become more complex.
Aside from a small overhead, the total memory consumption is very close to the sum of the memory used to represent each of these three matrices.

As a final note, \FFSol{} is not intended for very small systems since the backend needs to be recompiled for each prime.
The overhead from calling the compiler is independent of the matrix size so this only negatively impacts small matrices, which is perfectly acceptable since these matrices are already trivial to solve.
Once the time to solve the system overshadows the compile time then \FFSol{} becomes more performant than Mathematica's built in \code{Solve} or \code{Reduce}.
Depending on the computer, the tipping point occurs in the high hundreds to low thousands of parameters when the time to solve the matrix takes of order one second.

\section{User guide}
\label{sec:guide}

This section describes how to install \FFSol{} along with the major functions included in the package.
The package is available at
\begin{center}
\url{https://github.com/jfmangan/FiniteFieldSolve}.
\end{center}
The repository contains four relevant files: InstallScript.m, Examples.m, FiniteFieldSolve.m, and RowReduceLink.cpp.
Rather than cloning the repository with \code{git}, it is recommended to download files individually as described below.
Download Examples.m and InstallScript.m and place them in any desired directory.
As the name suggests, Examples.m provides several examples that can be run once \FFSol{} has been properly installed.
One of the included examples is discussed in Section \ref{sec:example}.
To install \FFSol{} run InstallScript.m from any directory.
This script will create a folder called FiniteFieldSolve in \$UserBaseDirectory/Applications and then it will download FiniteFieldSolve.m and RowReduceLink.cpp and place them inside the new directory.\footnote{On startup, FiniteFieldSolve.m will automatically determine its location in the filesystem, so it is possible for users to install the package outside of \$UserBaseDirectory.  However, it is still recommended to use InstallScript.m as this will ensure that Mathematica can load the package.}
Running \FFSol{} requires a C++ compiler.
Even though \FFSol{} automatically detects the operating system, setting up the compiler varies between machines.
\begin{itemize}
\item \emph{Windows} is not natively supported.
In order to run \FFSol{} on Windows it is suggested to use Windows Subsystem for Linux (WSL) \cite{WSL}, Cygwin \cite{cygwin}, or MinGW \cite{mingw}, at which point the Windows installation becomes the same as for Linux.
\FFSol{} has been tested with WSL2.
\item \emph{Linux} ships with gcc so when \FFSol{} detects Linux as the operating system, this is the compiler it will use.
\item \emph{MacOS} does not ship with a compiler so one must be installed.
The simplest option is to use clang.
If it is not already installed, it can be installed by running
\begin{center}
\code{xcode-select --install}
\end{center}
in a terminal.
By default, \FFSol{} will use clang on Mac.
Using a different compiler, like gcc, requires modifying the `gccString' in FiniteFieldSolve.m.
\FFSol{} has been tested on Macs with both Intel and Apple silicon processors.
\end{itemize}
Although OpenMP is recommended, it is disabled by default because its proper configuration is beyond the scope of this installation guide.
If OpenMP is installed and the compiler and environment variables are configured properly, OpenMP can be enabled by changing the `IsOpenMPInstalled' flag in FiniteFieldSolve.m from False to True.
The speed improvements from OpenMP are discussed in Section \ref{sec:example}.
Once \FFSol{} has been installed, it can be loaded in Mathematica with
\begin{center}
\code{\textless \textless FiniteFieldSolve\`{}}
\end{center}
just like any other package.
With installation complete, it is time to discuss the three most important high-level functions in the package.

\code{FiniteFieldSolve[equations, (optional: OptionsList)]} takes a list of equations and returns the solution as a list of replacement rules.
For example,
\begin{center}
\code{FiniteFieldSolve[\{a==b, a==1\}]}
\end{center}
returns \code{\{a$\rightarrow$1, b$\rightarrow$1\}}.
\FFSol{} will automatically detect the variables in the equations.
If an equation does not contain `\code{==}', then it is assumed that the equation is equal to zero.
\OL{} is an optional parameter of the form \code{\{string1, string2...\}} where the strings are treated as case independent.
\FFSol{} will print more output if \code{`verbose'} is included in \OL{}.
By default the equations are sorted by their density before they are solved.
To disable this feature include \code{`NoRowSorting'} in \OL{}.
By default \FFSol{} will use the first prime it solves over to determine the linearly independent rows and only use those rows when solving over subsequent primes.
This behavior can be disabled by including \code{`KeepLinearDepRows'} in \OL{}.
By default \FFSol{} will use the first prime to detect if the equations set any variables to zero.
These variables will not be solved for in the future and the ensuing linearly dependent rows will be removed by row reducing over the second prime.
To disable this behavior include \code{`KeepZeroVariables'} in \OL{}.
Row reduction will default to 16-bit primes which are generally faster and use less memory.
To use 32-bit primes include \code{`32bit'} in \OL{}.
For dense equations -- such as those generated by random numerical sampling over the rationals -- it can be useful to clear denominators to avoid $1/p$ factors.
To clear all denominators add \code{`ClearDenominators'} to \OL{}.
By default, \FFSol{} statically allocates a contiguous block of memory for the matrix to improve performance.  This may not be possible for sufficiently large matrices, in which case the matrix can by dynamically allocated by adding \code{`dynamic'} to \OL{}.

\code{FindLinearlyIndependentEquations[equations, (optional: prime), (optional: OptionsList)]} takes a list of equations and returns the linearly independent ones.
The advantage of this function is that it is typically faster than solving the whole system.
When building up a complicated system of equations, it can be helpful to periodically prune unnecessary equations with this function.
Failing to do so can result in a final system that is vastly overcomplete, increasing the time and (peak) memory needed to solve it.
This function also gives an accurate estimate of the rank of a system.
If equations are being generated by numerical sampling, then the number of variables minus the rank gives an upper bound on the number of equations that must be generated.
By default the function uses the prime 65,521, the largest 16-bit prime, but a different prime can be specified.
This function also defaults to sorting the equations by density before row reduction so that the function returns the sparsest linearly independent equations.
This behavior can be disabled by including \code{`NoRowSorting'} in \OL{}.
By default, \code{Find\-Linearly\-Independent\-Equations} statically allocates a contiguous block of memory for the matrix to improve performance.
This may not be possible for sufficiently large matrices, in which case the matrix can by dynamically allocated by adding \code{`dynamic'} to \OL{}.

\code{ConsistentEquationsQ[equations, (optional: prime), (optional: OptionsList)]} takes an inhomogeneous system of equations and returns False if the system is inconsistent and True otherwise.
This function operates by row reducing the system over a given prime, which defaults to 65,521, and then checking if the row reduction is inconsistent.
If the equations are inconsistent over the prime, then they are guaranteed to be inconsistent over the rationals.
To detect an inconsistency, it is faster to run this function than to reconstruct the full rational solution since the former only requires a single prime but the latter typically requires several.
For this reason, this function can be a useful diagnostic tool when an inconsistent system is expected.
By default, \code{Consistent\-EquationsQ} statically allocates a contiguous block of memory for the matrix to improve performance.
This may not be possible for sufficiently large matrices, in which case the matrix can by dynamically allocated by adding \code{`dynamic'} to \OL{}.

The three functions described above share certain features in common.
They all internally convert the equations to a matrix (in the form of a \code{SparseArray}) by calling \code{CoefficientArrays}.
The matrix may take up less memory than the equations themselves, so for advanced usage users may wish to work directly with the matrix.
In this case, the three relevant functions are \code{FiniteFieldSolveMatrix}, \code{FindLinearlyIndependentRows}, and \code{ConsistentMatrixQ}.
The precise documentation for these functions can be found in FiniteFieldSolve.m.

The workhorse behind all of the functions mentioned above is \code{RowReduceOverPrime} which takes a matrix and row reduces it over a given prime.
Critically, the rows are never rearranged during row reduction, rows of zeros are never deleted, and the first available pivot is always selected when scanning the matrix from left to right and top to bottom.
This means that \code{RowReduceOverPrime} does not return a matrix in strict row echelon form.
However, this algorithm has the advantage that its output can be used to trivially identify the linearly independent rows in the input matrix.\footnote{Specifically, if row $i$ in the row reduced matrix is a row of zeros, then row $i$ in the original matrix represents a linearly dependent equation.}
Furthermore, these rows will come from the top of the matrix where \FFSol{} stores the sparsest equations.
So when solving a full system over multiple primes \FFSol{} can use \code{RowReduceOverPrime} to find the linearly independent, sparsest equations over the first prime to speed up row reduction over later primes.
The arguments for \code{RowReduceOverPrime} are given below.

\code{RowReduceOverPrime[matrix, prime, (optional: StaticOrDynamicMem), (optional: RowsToUse), (optional: ColsToUse)]} row reduces a \code{matrix} over the \code{prime}.
The \code{prime} must be less than $2^{32}$ where a \code{prime} less than $2^{16}$ automatically switches the function into its faster and more memory efficient 16-bit mode.
As described above, rows are not rearranged during reduction so the output matrix may contain rows of zeros and may not be in strict row echelon form.
\code{Row\-Reduce\-Over\-Prime} will return \code{\$Failed} if the matrix could not be projected over the given \code{prime}.
\code{Rows\-To\-Use} is an optional argument that defaults to \code{All}.
If only certain rows of \code{matrix} should be used or the rows should be rearranged, this can be accomplished by setting \code{Rows\-To\-Use} to the desired list of rows, for example \{4, 1, 5,...\}.
\code{Rows\-To\-Use} enables \FFSol{} to sort the rows of a matrix by their density before solving it.
Storing sparser equations above denser ones reduces fill-in thereby improving performance.
\code{Cols\-To\-Use} is the analog of \code{Rows\-To\-Use} but for columns.
\code{Static\-Or\-Dynamic\-Mem} is an optional argument that controls how the memory for the \code{matrix} is allocated.
\code{Static\-Or\-Dynamic\-Mem} defaults to \code{`static'} meaning that the \code{matrix} will be statically allocated as a contiguous block of memory to improve access times.
For sufficiently large matrices there may not be a contiguous block of memory available so the \code{matrix} may need to be dynamically allocated by setting \code{Static\-Or\-Dynamic\-Mem} to \code{`dynamic'}.

\section{Worked example and benchmarks}
\label{sec:example}

This section describes two applications of \FFSol{} to bootstrapping relativistic scattering amplitudes.
The first example shows how to determine the 8pt tree amplitude for Dirac-Born-Infeld (DBI) theory using its soft behavior \cite{Cheung:2014dqa}.
This calculation is worked through in full detail in the example file included with \FFSol{}.
The second example is an overview of determining the 4pt two-loop color-dual integrand for the non-linear sigma model (NLSM) \cite{Edison:2023ulf}.
\FFSol{} is benchmarked against other solvers in these examples.
While the two examples in this section are drawn from real-world physics problems, they only involve moderately dense systems.
Since \FFSol{} is designed to handle fully dense systems, one is benchmarked in \ref{sec:appendixDense}.
The dense example is not presented here because it lacks the same level of physical motivation.

\subsection{Soft bootstrap of 8pt DBI}
\label{sec:DBIExample}

DBI is a massless scalar field theory that is uniquely determined by its infrared behavior \cite{Cheung:2014dqa}.
With a mostly plus metric signature and in units where the coupling constant is unity, the DBI Lagrangian is
\begin{align}
\label{eq:DBILagrangian}
\mathcal{L} = - \sqrt{1+(\partial \phi)^2} \, .
\end{align}
In order to obtain a non-trivial linear system, we will bootstrap the 8pt tree amplitude $A_8$ assuming that the 4pt and 6pt amplitudes have already been computed using the bootstrap.
The amplitude is built from three separate types of graphs, two of which have factorization channels and the third represents a contact term
\begin{align}
\label{eq:DBIGraphs}
{
\begin{tikzpicture}
\begin{feynman}
\vertex (v0a) at (-1,0);
\vertex (v1a) at (0,0);
\vertex (v2a) at (1,0);
\vertex (p1a) at (-1.707,-0.707);
\vertex (p2a) at (-2,0);
\vertex (p3a) at (-1.707,0.707);
\vertex (p4a) at (0,1);
\vertex (p5a) at (1.707,0.707);
\vertex (p6a) at (2,0);
\vertex (p7a) at (1.707,-0.707);
\vertex (p8a) at (0,-1);
\vertex (v0b) at (3.5+0,0);
\vertex (v1b) at (3.5+1,0);
\vertex (p1b) at (3.5+0,-1);
\vertex (p2b) at (3.5-0.707,-0.707);
\vertex (p3b) at (3.5-1,0);
\vertex (p4b) at (3.5-0.707,0.707);
\vertex (p5b) at (3.5+0,1);
\vertex (p6b) at (3.5+1.707,0.707);
\vertex (p7b) at (3.5+2,0);
\vertex (p8b) at (3.5+1.707,-0.707);
\vertex (v0c) at (7+0,0);
\vertex (p1c) at (7-1,0){};
\vertex (p2c) at (7-0.707,0.707){};
\vertex (p3c) at (7+0,1){};
\vertex (p4c) at (7+0.707,0.707){};
\vertex (p5c) at (7+1,0){};
\vertex (p6c) at (7+0.707,-0.707){};
\vertex (p7c) at (7+0,-1){};
\vertex (p8c) at (7-0.707,-0.707){};
\diagram{
(v0a) -- [ultra thick,](v1a),
(v1a) -- [ultra thick,](v2a),
(v0a) -- [ultra thick,](p1a),
(v0a) -- [ultra thick,](p2a),
(v0a) -- [ultra thick,](p3a),
(v1a) -- [ultra thick,](p4a),
(v2a) -- [ultra thick,](p5a),
(v2a) -- [ultra thick,](p6a),
(v2a) -- [ultra thick,](p7a),
(v1a) -- [ultra thick,](p8a),
(v0b) -- [ultra thick,](v1b),
(v0b) -- [ultra thick,](p1b),
(v0b) -- [ultra thick,](p2b),
(v0b) -- [ultra thick,](p3b),
(v0b) -- [ultra thick,](p4b),
(v0b) -- [ultra thick,](p5b),
(v1b) -- [ultra thick,](p6b),
(v1b) -- [ultra thick,](p7b),
(v1b) -- [ultra thick,](p8b),
(v0c) -- [ultra thick,](p1c),
(v0c) -- [ultra thick,](p2c),
(v0c) -- [ultra thick,](p3c),
(v0c) -- [ultra thick,](p4c),
(v0c) -- [ultra thick,](p5c),
(v0c) -- [ultra thick,](p6c),
(v0c) -- [ultra thick,](p7c),
(v0c) -- [ultra thick,](p8c),
};
\end{feynman}
\end{tikzpicture}
}
\, .
\end{align}
The two graphs on the left can be obtained by re-using the amplitudes obtained from earlier stages in the bootstrap, leaving only the contact diagram to be determined.
From the DBI Lagrangian, the contact term should be proportional to
\begin{align}
(p_1 p_2) (p_3 p_4) (p_5 p_6) (p_7 p_8) + \text{perms}
\end{align}
where $p_i$ is the momentum of particle $i$ and the sum extends over all permutations of external labels.
However, we will confirm by explicit calculation that the contact term is uniquely fixed by the soft behavior of DBI without any reference to the Lagrangian.

The ansatz for the contact term is constructed as follows.
There are 20 generalized Mandelstam invariants $p_i \cdot p_j$ that are independent on-shell, that is, after imposing
\begin{align}
\left\{p_i^2=0 ~|~ 1 \leq i \leq n \right \} \text{ and } \sum \limits_{i=1}^n p_i =0 \, .
\end{align}
The exact basis chosen does not matter.
The basis in the example file includes $(p_2 p_4)$ and $(p_2 p_5)$ as well as 18 more dot products.
There are 8,855 independent terms with mass dimension $p^8$ so the contact term looks schematically like
\begin{align}
A_{8, \,\text{contact}} (p_1, p_2,... p_8) = c_1 (p_2 p_4)^4 + c_2 (p_2 p_4)^3 (p_2 p_5)+...
\end{align}
The first constraint on this ansatz is that the contact term should be Bose symmetric, that is, the contact term should be invariant under any relabeling of the legs.
Full permutation invariance can be guaranteed by imposing two independent symmetry relations.
First, the contact term should be invariant under cyclically rotating all of the labels by one position,
\begin{align}
\label{eq:cyclic}
A_{8, \,\text{contact}} (p_1, p_2, p_3, p_4, p_5, p_6, p_7, p_8) = A_{8, \,\text{contact}} (p_2, p_3, p_4, p_5, p_6, p_7, p_8, p_1),
\end{align}
and, second, the contact term should be invariant under swapping particles 1 and 2,
\begin{align}
\label{eq:swap}
A_{8, \,\text{contact}} (p_1, p_2, p_3, p_4, p_5, p_6, p_7, p_8) = A_{8, \,\text{contact}} (p_2, p_1, p_3, p_4, p_5, p_6, p_7, p_8).
\end{align}

Naively each of the conditions in \eqref{eq:cyclic} and \eqref{eq:swap} produces 8,855 equations -- as many equations as there are terms in the ansatz -- but some of the equations become trivial in the right Mandelstam basis.
Table \ref{tbl:benchmark1} documents how long it takes \FFSol{} to solve the linear system along with the dimensions and density of the system.
The table also contains timings against Mathematica's \code{Solve}, \LinSysSol{} \cite{LinSysSol}, \spasmlink{} \cite{spasmlink}, \kira{} \code{2} \cite{Klappert:2020nbg} (using both \firefly{} \cite{Klappert:2019emp} and \fermat{} \cite{fermat})\footnote{The author would like to thank a referee for elucidating the various options within \kira{}.}, and \FiniteFlow{} \cite{Peraro:2019svx} (using both its sparse and dense solvers).
The testing was done on two different machines, one against \spasmlink{} and one against \kira{}.
When comparing to \spasmlink{}, \kira{}, and the sparse solver in \FiniteFlow{} it is worth remembering that these packages are intended for potentially very sparse systems.
\kira{} does not directly interface with Mathematica but the time to convert between the two programs was not included in the timings in the table.
In the benchmarks presented in Table \ref{tbl:benchmark1}, OpenMP was enabled for \FFSol{} on both machines.
Disabling OpenMP has virtually no effect on overall performance in this instance, but this will not be the case for the next example.
By declaring the prime in the C++ backend as `\code{volatile}' it is possible to estimate the performance that would be lost if \FFSol{} did not compile its primes.
Failing to compile the primes results in the overall solve time increasing by roughly a factor of 2.
The time savings from compiling the prime will be more dramatic in the next example.
As the table demonstrates, \FFSol{} is about two orders of magnitude faster than Mathematica's \code{Solve} and uses about an order of magnitude less memory.
In this test, \FFSol{} is the fastest solver benchmarked but certain configurations of other solvers are more memory efficient.
\FiniteFlow{}'s sparse solver consumes roughly 10\% less memory than \FFSol{} while being about 6 times slower.
\kira{} with \fermat{} is the most memory efficient option, using only 22\% as much memory as \FFSol{} at the cost of being 3.6 times slower.
A breakdown of where \FFSol{} spends most of its time is given in Table \ref{tbl:breakdown1}.
As might be expected, row reduction takes up the majority of the time.

\begin{table}
\centering
\caption{\FFSol{} was benchmarked against other solvers in bootstrapping the 8pt DBI contact term.
Note that \spasmlink{} and \kira{} are intended for very sparse systems.  The ratios in the table are with respect to the time and memory it takes \FFSol{} to solve the system.}
\subcaption*{Benchmarks on MacOS with i5-7267U, 16GB memory, Mathematica 13.2, and GCC 12.2}
\begin{tabular}{l |c|c|c|c}
	& Time (s) & Time ratio & Memory (GB) & Memory ratio\\
\hline
\FFSol{} & 11 & 1 & 0.50 & 1\\
Mathematica \code{Solve} & 1400 & 130 & 9.7 & 19\\
\LinSysSol{} & 2200 & 200 & 5.6 & 11\\
\spasmlink{} & 1200 & 110 & 0.96 & 1.9\\
\begin{tabular}{@{}l@{}} \FiniteFlow{} \\  ~~~~ \ffdense{} \end{tabular} & 780 & 71 & 8.9 & 18 \\
\begin{tabular}{@{}l@{}} \FiniteFlow{} \\  ~~~~ \ffsparse{} \end{tabular} & 65 & 5.9 & 0.49 & 0.98
\end{tabular}
\bigskip
\subcaption*{Benchmarks on Ubuntu under WSL2 with i5-6600, 32GB memory, Mathematica 13.1, and GCC 9.4}
\begin{tabular}{l |c|c|c|c}
	& Time (s) & Time ratio & Memory (GB) & Memory ratio\\
\hline
\FFSol{} & 8.5 & 1 & 0.52 & 1\\
Mathematica \code{Solve} & 1500 & 180 & 24 & 46\\
\LinSysSol{} & 1900 & 220 & 5.7 & 11\\
\kira{} \code{2} (\firefly{}) & 180 & 21 & 0.21 & 0.40\\
\kira{} \code{2} (\fermat{}) & 31 & 3.6 & 0.11 & 0.22\\
\begin{tabular}{@{}l@{}} \FiniteFlow{} \\  ~~~~ \ffdense{} \end{tabular} & 660 & 76 & 14 & 27 \\
\begin{tabular}{@{}l@{}} \FiniteFlow{} \\  ~~~~ \ffsparse{} \end{tabular} & 55 & 6.5 & 0.46 & 0.88
\end{tabular}
\bigskip
\subcaption*{Matrix properties}
\begin{tabular}{cc}
Matrix dimensions & 16,705 $\times$ 8,855 \\
Matrix density & $3.4 \times 10^{-3}$\\
Highest row density & $2.4 \times 10^{-2}$
\end{tabular}
\label{tbl:benchmark1}
\end{table}

\begin{table}
\centering
\caption{
The table below gives the breakdown of where \FFSol{} spends most of its time when solving the system in Table \ref{tbl:benchmark1}.
The breakdown is virtually identical between machines.
Unlike subsequent benchmarks, only a single row reduction is performed so the Chinese remainder theorem is never used.
Even though the matrix is only row reduced once, it is projected over an additional prime to check the putative solution.
In general, if the matrix is row reduced $n$ times, then it will be projected over $n+1$ primes.
}
Number of row reductions used to solve the system:  1

\vspace*{1em}

\begin{tabular}{l  l}
Time breakdown: & \\
\hline
Row reduction & 80\% \\
Extended Euclidean algorithm & 10\% \\
Chinese remainder theorem & 0\% \\
Converting equations to matrix & 10\%
\end{tabular}
\label{tbl:breakdown1}
\end{table}

After imposing Bose symmetry there are only 5 free parameters left in the contact term.
All remaining freedom in the ansatz can be fixed by constructing $A_8$ from \eqref{eq:DBIGraphs} and imposing the soft theorem of DBI.\footnote{There are other principles that fully fix the amplitude including classical conformal invariance \cite{Cheung:2020qxc}.}
Amplitudes in this theory vanish as momentum squared in the soft limit, that is,
\begin{align}
\lim \limits_{z\to 0} A_n(z p_i) = \mathcal{O}\left( z^2 \right)
\end{align}
where particle $i$ is taken soft.
At a practical level, when taking the soft limit of an on-shell amplitude, it is key that $p_i$ does not play any special role in enforcing the on-shell conditions.
For example, if momentum conservation was used to eliminate $p_n$ from $A_n$, this does not mean that $A_n$ is identically zero in the soft limit of $p_n$.
After scaling the momentum of one of the unprivileged particles via $p_i \to z p_i$, there are multiple methods to extract the actual equations enforcing the soft limit.
One option would be to call \code{Together} on the entire amplitude.
A more appropriate method in this case is to sample different kinematical configurations numerically by repeatedly mapping the Mandelstams $(p_j p_k)$ to random integers.
Each choice of random Mandelstams will generally generate a different fully dense equation enforcing the soft limit.
One of the convenient features of \FFSol{} is that it can easily handle these dense equations, sidestepping an expensive call to \code{Together}.
However, since there are only 5 remaining parameters, Mathematica's \code{Solve} is more than sufficient to complete the bootstrap.
After imposing the soft limit, the amplitude is fully fixed and the contact term corresponds to the 8pt interaction in \eqref{eq:DBILagrangian}.

\subsection{Color-dual bootstrap of 4pt two-loop NLSM}

The next example is the construction of color-dual 4pt two-loop numerators for the non-linear sigma model (NLSM).
This example is more complex than the previous one, but a brief overview is provided below.
For the full calculation see Ref \cite{Edison:2023ulf}.
The benchmarks in Table \ref{tbl:benchmark2} can be understood without any knowledge of how the system was constructed.
The following discussion assumes some familiarity with color-kinematics duality and the double copy \cite{Bern:2010ue, Bern:2008qj}.
Recent reviews of these topics include Refs \cite{Bern:2019prr, Bern:2022wqg, Adamo:2022dcm}.

Color-kinematics duality applies to certain special theories, including Yang-Mills and NLSM, that carry color (or flavor) indices.
The $n$-pt $L$-loop scattering amplitude in any colored theory can be written as
\begin{align}
\label{eq:nptLloopAmp}
\mathcal{A}_n^L = \sum \limits_\Gamma \frac{1}{S_\Gamma} \int \left( \prod \limits_{i=1}^L \frac{d^D \ell_i}{(2\pi)^D}\right) \frac{C_\Gamma N_\Gamma}{d_\Gamma}
\end{align}
where $D$ is the spacetime dimension and the sum runs over \emph{cubic} graphs $\Gamma$.
The amplitude can always be written as a sum over purely cubic graphs by multiplying and dividing by propagators.
Here $S_\Gamma$ is the symmetry factor associated with the graph $\Gamma$, $d_\Gamma$ is the product of propagators, $C_\Gamma$ is the color factor made of a product of structure constants $f^{abc}$, and $N_\Gamma$ is the kinematic numerator made of dot products of momentum and polarization vectors.
The color factors in \eqref{eq:nptLloopAmp} obey a set of Jacobi identities.
Color-kinematics duality states that there exists a way to write the kinematic numerators so that they obey the same Jacobi relations as the color factors.
Only a few Lagrangians are known to manifest color-kinematics duality \cite{Edison:2023ulf, Monteiro:2011pc, Ben-Shahar:2021zww, Cheung:2020djz, Cheung:2021zvb, Cheung:2022mix, Cheung:2016prv, Ben-Shahar:2021doh, Ben-Shahar:2022ixa}, so color-dual numerators are typically determined using an ansatz.
Once color-dual numerators are found, the amplitude for the ``double-copy'' theory is obtained by replacing $C_\Gamma$ with another copy of $N_\Gamma$ in \eqref{eq:nptLloopAmp}.

As a brief example of color-kinematics duality, consider the NLSM 4pt tree amplitude.\footnote{Tree-level amplitudes in NLSM are fully fixed by color-kinematics duality \cite{Carrasco:2019qwr}.}
The only graph topology contributing to this process is
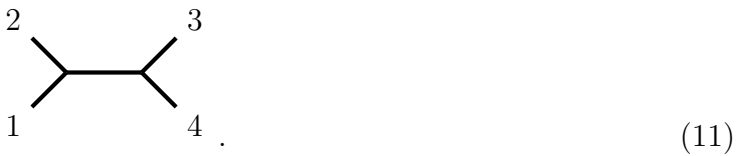
\begin{align}
{
\begin{tikzpicture}
\begin{feynman}
\vertex(v0) at (0,0);
\vertex(v1) at (1,0);
\vertex(p1) at (-0.707,-0.707){1};
\vertex(p2) at (-0.707,0.707){2};
\vertex(p3) at (1+0.707,0.707){3};
\vertex(p4) at (1+0.707,-0.707){4};
\diagram{
(v0) -- [ultra thick,](v1),
(v0) -- [ultra thick,](p1),
(v0) -- [ultra thick,](p2),
(v1) -- [ultra thick,](p3),
(v1) -- [ultra thick,](p4),
};
\end{feynman}
\end{tikzpicture}
}
\, .
\end{align}
The color factor associated with this ($s$-channel) graph is $C_{(12|34)} = f^{12c}f^{c34}$.
The color-dual form of the pion numerator can be found by making an ansatz with the appropriate power counting,
\begin{align}
\label{eq:Ansatz}
N_{(12|34)} = c_1 s^2 + c_2 s t + c_3 t^2,
\end{align}
where $s=(p_1+p_2)^2$, $t=(p_2 +p_3)^2$, and $u=-s-t$.
This ansatz implicitly assumes that the numerator is a local function, which is not absolutely necessary for color-kinematics duality and can be an overly restrictive assumption at times \cite{Edison:2023ulf, Cheung:2021zvb, Brandhuber:2021bsf, Bern:2015ooa, Mogull:2015adi}.
Aside from locality and power counting, three explicit constraints are imposed on the ansatz.
First, the numerator must respect all of the symmetries of its associated graph.
At 4pt tree level the graph symmetries are spanned by
\begin{align}
\label{eq:GraphSym}
N_{(12|34)} &= - N_{(21|34)}\\
N_{(12|34)} &= N_{(34|12)},
\end{align}
where the sign compensates for the signs buried in the structure constants making up the color factor.
Strictly speaking color-kinematics duality does not require that numerators respect their graph symmetries, but it is a convenient and natural requirement just like postulating local ansatz\"{e} for numerators.
Another benefit of enforcing graph symmetries is that the $t$- and $u$-channel numerators can be obtained by relabeling the $s$-channel numerator $N_{(12|34)}$.
The second constraint is that the kinematic numerators should obey the same Jacobi relations as the color factors.
In this case, the color factors obey
\begin{align}
C_{(12|34)} + C_{(23|14)} + C_{(31|24)} =0
\end{align}
so the numerators are required to satisfy the same condition
\begin{align}
\label{eq:Jacobi}
N_{(12|34)} + N_{(23|14)} + N_{(31|24)} =0\, .
\end{align}
The third and final requirement is that the amplitude factorizes correctly on its poles, or in general, that the integrand has the correct unitarity cuts \cite{Bern:1994cg, Britto:2004nc, Bern:2011qt}.
For 4pt NLSM this is trivial since the on-shell 3pt amplitude vanishes on general grounds.
Together, \eqref{eq:GraphSym} and \eqref{eq:Jacobi} fix the ansatz in \eqref{eq:Ansatz} up to an overall coupling constant, giving
\begin{align}
N_{(12|34)} \propto s(t-u)\, .
\end{align}
The fully color-dressed pion amplitude is
\begin{align}
A_4 = \frac{C_{(12|34)} N_{(12|34)}}{(p_1+p_2)^2} + \text{cyclic}(1,2,3),
\end{align}
where the sum extends over cyclic permutations of labels 1, 2, and 3.
With the color-dual numerators in hand, the double copy produces the 4pt special Galileon amplitude
\begin{align}
A_4 = \frac{N^2_{(12|34)}}{(p_1+p_2)^2} + \text{cyclic}(1,2,3) \propto s t u \, .
\end{align}

For the case of the 4pt two-loop NLSM integrand constructed in Ref \cite{Edison:2023ulf}, power counting requires that the numerator scales as $p^{12}$.
There are 8,008 independent scalar invariants at this mass dimension and since two diagrams require ansatz\"{e}, the entire problem involves 16,016 parameters, at least for the case of a local numerator.\footnote{For technical reasons there are actually two more parameters in the system, one of which is used to represent inhomogeneous terms in equations as explained in Section \ref{sec:FFMethods}.}
The key constraints are the same as they were for the 4pt tree example, namely, the numerators must respect graph symmetries, the numerators must obey the same linear relations as the color factors, and the integrand must factorize correctly.
As a matter of convenience, the Jacobi relations are imposed first so that the basis of numerators can be reduced to just two graphs.
The Jacobi relations produce 59,544 equations, where the exact number depends on implementation details.
The linearly independent subset of 11,543 equations can be selected using \code{FindLinearlyIndependentEquations}.
Solving the equations is unnecessary at this stage and will actually overcomplicate the problem as this intermediary solution is more complex than the final solution.
Next, imposing graph symmetries produces 115,494 equations, which are similarly implementation-dependent.
Combining the graph symmetry equations with the Jacobi relations produces 14,773 linearly independent equations.
Yet again, it is unnecessary to actually solve the equations; finding the linearly independent subset is enough.

The third key constraint is that the integrand has the correct unitarity cuts.
While the Jacobi relations and the graph symmetries are polynomial constraints, the unitarity cuts generate constraints that are sums of rational functions where the denominators come from uncut momenta.
Extracting the actual equations requires combining fractions with something like \code{Together} in Mathematica.
While this is straightforward in theory, this can become a severe bottleneck in practice.
Alternatively, the equations can be extracted by plugging in random integers for the kinematics appearing in the rational functions.
Each choice of random kinematics yields another equation.
Since \code{FindLinearlyIndependentEquations} effectively computes the rank of the equations already generated, it is possible to accurately estimate the number of equations that need to be produced numerically.
In this case, 16,016 $ - $ 14,773 $ = $ 1,243 equations need to be generated by random sampling.
Note that it is trivial to parallelize numerical sampling but the same is not true for combining large rational functions.
The most important advantage of numerical sampling is that \FFSol{} allows the user to offload all of the complications onto the computer; no human effort needs to be put into finding or maintaining the sparse equations that come from combining rational functions.
The disadvantage of sampling is that, first, it produces dense equations and, second, it requires a method for computing the rank.
\FFSol{} addresses both issues.
The rank of the graph symmetry and Jacobi equations could be obtained by directly solving them, but this is both overkill and slow since the intermediary solution requires more primes to reconstruct than the final solution.
Aside from computing the rank, periodically culling the system of linearly dependent equations also has the advantage that it reduces the memory required to solve the final system.
After imposing unitarity cuts, the system has 15,251 equations.
Since there are 16,016 ansatz parameters, there is still room to enforce additional aesthetic constraints that simplify the answer.
The last constraints come from matching the NLSM numerators to those of Zakharov-Mikhailov theory in two spacetime dimensions \cite{Cheung:2022mix, Zakharov:1973pp}.

\begin{table}
\centering
\caption{\FFSol{} was benchmarked against other solvers in bootstrapping the 4pt two-loop NLSM integrand.
Note that \spasmlink{} and \kira{} are intended for very sparse systems.  The ratios in the table are with respect to the time and memory it takes \FFSol{} to solve the system.}
\subcaption*{Benchmarks on MacOS with i5-7267U, 16GB memory, Mathematica 13.2, and GCC 12.2}
\begin{tabular}{l |c|c|c|c}
	& Time (s) & Time ratio & Memory (GB) & Memory ratio\\
\hline
\FFSol{} & 380 & 1 & 0.81 & 1\\
\spasmlink{} & $1.6\times10^4$ & 42 & 4.4 & 5.4\\
\begin{tabular}{@{}l@{}} \FiniteFlow{} \\  ~~~~ \ffdense{} \end{tabular} & 1900 & 5.0 & 9.9 & 12 \\
\begin{tabular}{@{}l@{}} \FiniteFlow{} \\  ~~~~ \ffsparse{} \end{tabular} & 5300 & 14 & 6.1 & 7.5
\end{tabular}
\bigskip
\subcaption*{Benchmarks on Ubuntu under WSL2 with i5-6600, 32GB memory, Mathematica 13.1, and GCC 9.4}
\begin{tabular}{l |c|c|c|c}
	& Time (s) & Time ratio & Memory (GB) & Memory ratio\\
\hline
\FFSol{} & 420 & 1 & 1.0 & 1\\
\kira{} \code{2} (\firefly{}) & $2.4 \times 10^5$ & 570 & 2.6 & 2.6\\
\kira{} \code{2} (\fermat{}) & 4100 & 9.8 & 1.9 & 1.9\\
\begin{tabular}{@{}l@{}} \FiniteFlow{} \\  ~~~~ \ffdense{} \end{tabular} & 1800 & 4.3 & 25 & 25 \\
\begin{tabular}{@{}l@{}} \FiniteFlow{} \\  ~~~~ \ffsparse{} \end{tabular} & 4400 & 10 & 7.9 & 7.9
\end{tabular}
\bigskip
\subcaption*{Matrix properties}
\begin{tabular}{cc}
Matrix dimensions & 15,959 $\times$ 16,018 \\
Matrix density & $6.4 \times 10^{-2}$\\
Highest row density & 1
\end{tabular}
\label{tbl:benchmark2}
\end{table}

\begin{table}
\centering
\caption{
The table below gives the breakdown of where \FFSol{} spends most of its time when solving the system in Table \ref{tbl:benchmark2}.
The breakdown is virtually identical between machines.
}
Number of row reductions used to solve the system:  3

\vspace*{1em}

\begin{tabular}{l  l}
Time breakdown: & \\
\hline
Row reduction & 60\% \\
Extended Euclidean algorithm & 20\% \\
Chinese remainder theorem & 10\% \\
Converting equations to matrix & 10\%
\end{tabular}
\label{tbl:breakdown2}
\end{table}

Once all of the equations have been assembled, it is finally time to solve them.
Since the system was assembled with \FFSol{}, all of the linearly dependent equations have been removed, leaving only the essential equations left to solve.
\FFSol{} was timed against \spasmlink{}, \kira{}, and \FiniteFlow{} in solving the system and the results are presented in Table \ref{tbl:benchmark2}.
Again, recall that \spasmlink{}, \kira{}, and \FiniteFlow{}'s sparse routine are intended for (very) sparse systems, which is not the case for the fully dense rows generated by numerically sampling the unitarity constraints.
Even though the linear system has been timed on different solvers, \FFSol{} was instrumental in constructing it.
Similar to the DBI example, \FFSol{} is the fastest option tested, sometimes by an order of magnitude or more.
Unlike the DBI example, \FFSol{} is now the most memory efficient option as well.
\kira{} with \fermat{} is the second most memory efficient option, consuming roughly twice as much memory as \FFSol{} while being about ten times slower.
OpenMP was enabled for \FFSol{} when constructing the results in Table \ref{tbl:benchmark2}.
If OpenMP is disabled, the solve times increase by roughly a factor of 1.4 on both machines, which is reasonable since both computers have dual-core CPUs.
More importantly, if the prime is not compiled, then overall performance drops by roughly a factor of 4 on both computers.
Since the NLSM system contains approximately twice as many parameters as the DBI system, it might be reasonable to expect that it would take very roughly $2^3$ as much time to solve.
The NLSM system takes longer to solve than expected because it is denser so \FFSol{} cannot skip as many rows during row reduction.
Furthermore, as shown in Table \ref{tbl:breakdown2}, the NLSM system requires three row reductions instead of the single row reduction needed for the DBI example.
As compared to the DBI example in Table \ref{tbl:breakdown1}, the weight has shifted away from row reduction towards the extended Euclidean algorithm and the Chinese remainder theorem.
There are two reasons for this.
First, the rational solution to the NLSM problem is more complex as indicated by the increased number of primes required for reconstruction.
Second, as mentioned in Section \ref{sec:technical}, subsequent row reductions are typically faster than the initial one, decreasing the overall proportion of time spent on row reduction.

\section{Discussion}
\label{sec:conclusions}

\FFSol{} is a general-purpose solver for exact linear systems over the rationals.
Systems of equations of this type appear prominently in high energy theory in integral reduction routines and amplitude bootstraps.
A worked example of a bootstrap problem was provided in the text and accompanying repository.
\FFSol{} uses well-known finite field methods to avoid intermediary expression swell and roundoff errors that might otherwise plague a traditional solver.
In testing, the new package is roughly two orders of magnitude faster and uses an order of magnitude less memory than Mathematica, opening the door to new physics problems.

A variety of techniques are used to improve performance in \FFSol{}.
First, the C++ backend is recompiled every time it is called.
Since the matrix dimensions are known at compile time, the memory can be statically allocated, improving memory access.
This technique is confined to dense solvers where memory for new nonzero values never needs to be allocated.
Compiling the backend also converts the ubiquitous and expensive modulo operations into a much faster combination of instructions.
Compiling the prime leads to overall speed increasing by up to a factor of 4 in testing.
Generally, any piece of software that performs a lot of modular arithmetic with respect to a fixed prime may benefit from re-compiling for each prime.
The actual time savings will depend on the proportion of time that the software spends on arithmetic.
The backend also improves memory usage and speed by using 16-bit primes, where 32-bit and 8-bit alternatives are slower in testing.
Due to its simple row reduction algorithm, the package benefits from two layers of parallelization.
Moving to the frontend, it improves performance by extracting as much information as possible over the first few primes, including finding the linearly independent equations of the system.
This can sometimes dramatically simplify the system that must be solved over subsequent primes.
The ability to easily identify linearly independent rows in conjunction with the solver's ability to efficiently manipulate dense equations frees the user to generate equations by any and all means, including numerical sampling.

Although some of the techniques in \FFSol{} may be applicable to other solvers, \FFSol{} is most directly suited to dense systems including those that appear in certain amplitude bootstraps.
One of the most promising applications of \FFSol{} is in bootstrapping higher-dimension color-dual operators since this can involve large dense systems, especially for scalar theories \cite{Carrasco:2022lbm, Carrasco:2022sck, Carrasco:2023wib, Chen:2023dcx, Chi:2021mio, Brown:2023srz, Li:2023wdm, Carrasco:2021ptp, Carrasco:2019yyn, Broedel:2012rc, Bonnefoy:2021qgu, Carrasco:2016ygv, Carrasco:2016ldy, Low:2020ubn, Low:2019wuv}.
Other promising applications may include:  bootstrapping higher multiplicity AdS amplitudes \cite{Alday:2023kfm, Alday:2023mvu, Alday:2023jdk, Alday:2022xwz, Alday:2022uxp, Alday:2021odx}; bootstrapping amplitudes from their physical properties like soft behavior, gauge invariance, factorization, conformal invariance, etc., \cite{Cheung:2014dqa, Cheung:2016drk, Cheung:2020qxc, Nutzi:2019ufl, Loebbert:2018xce, Cheung:2018oki, Ren:2022sws, Rodina:2018pcb, Arkani-Hamed:2016rak, Low:2019ynd}; and enumeration of higher-dimension or higher-multiplicity effective field theory operators \cite{Low:2022iim}.

Last, and perhaps most importantly, \FFSol{} is simple to use and straightforward to install.
\FFSol{} and Mathematica's \code{Reduce} and \code{Solve} share similar syntax so it is trivial to swap one out for the other.
Installing \FFSol{} on Linux amounts to running the installer script included in the repository.
The procedure is identical on MacOS except that the user may need to install a compiler by running a single command in the terminal.
\FFSol{} has been tested under Windows, but the installation process is more involved.

\paragraph{Acknowledgments}
The author would like to thank JJ Carrasco, Sasank Chava, Alex Edison, Kezhu Guo, Nic Pavao, and Laurentiu Rodina for feedback on the manuscript and software, insightful conversations, and related collaboration.
This work was supported by the DOE under contract DE-SC0015910 and by the Alfred P. Sloan Foundation.
J.M. additionally acknowledges the Northwestern University Amplitudes and Insight group, the Department of Physics and Astronomy, and Weinberg College for their generous support.
Feynman diagrams in this paper were typeset using TikZ-Feynman \cite{Ellis:2016jkw}.



\appendix

\section{Number theoretic algorithms}
\label{sec:appendix}

This appendix provides a brief summary of the number theoretic algorithms mentioned in the text including the extended Euclidean algorithm, Fermat's little theorem, and the Chinese remainder theorem.

Given integers $a$ and $b$, the Euclidean algorithm is a method for computing integers $x$ and $y$ such that $ax+by=(a,b)$, where $(a,b)$ denotes the greatest common divisor of $a$ and $b$.
The extended variant of the Euclidean algorithm starts by setting
\begin{alignat}{2}
& r_0 = a   \quad\quad\quad && r_1=b \nonumber \\
& s_0= 1   \quad\quad\quad && s_1=0\\
& t_0=0   \quad\quad\quad && t_1=1.\nonumber
\end{alignat}
The algorithm proceeds by repeated integer division
\begin{align}
&q_i = \left \lfloor \frac{r_{i-1}}{r_i} \right \rfloor \\
& r_{i+1} = r_{i-1} - q_i r_i \\
& s_{i+1} = s_{i-1} - q_i s_i\\
& t_{i+1} = t_{i-1} - q_i t_i
\end{align}
and terminates when $r_{n+1}=0$.
The values of $x$ and $y$ are given by $s_n$ and $t_n$, respectively.
Since $r_n=(a,b)$, the algorithm can be seen as a method for determining the greatest common divisor.
Furthermore, the extended Euclidean algorithm can be used for rational reconstruction of $a$ over $\mathbb{Z}_b$.
By induction, $as_i+bt_i = r_i$ so that $r_i/s_i$ is a \emph{potential} rational reconstruction of $a$.
To find \emph{the} rational reconstruction of $a$, the algorithm is stopped when $r_i < \sqrt{b/2}$.
If $s_i < \sqrt{b/2}$, then $r_i/s_i$ is the unique rational reconstruction of $a$ with numerator and denominator less than $\sqrt{b/2}$ \cite{RR1, RR2}.
Here $b$ corresponds to the product of primes $p_1 p_2...$ in Algorithm \ref{alg:RowReduce}.

The Euclidean algorithm can also be used to determine multiplicative inverses over $\mathbb{Z}_b$.
Assuming $a$ and $b$ are co-prime, the multiplicative inverse of $a$ is just $x$.
Another option for determining multiplicative inverses is to use Fermat's little theorem which states that $a^p \equiv a \mod p$ for $p$ prime.
In other words $a^{p-2}$ is the multiplicative inverse of $a$ assuming that $a$ is not a multiple of $p$.
The exact method used to determine inverses is largely irrelevant since row reducing an $n\times n$ matrix takes $\mathcal{O}(n^3)$ time overall but only requires computing at most $n$ inverses.

The final number theoretic result used in \FFSol{} is the Chinese remainder theorem.
Assuming $m_1$, $m_2$... are pairwise co-prime, the system of congruences
\begin{align}
&x \equiv c_1 \mod{m_1}\nonumber\\
&x \equiv c_2 \mod{m_2}\\
&... \nonumber
\end{align}
has a unique solution $x \mod m$ where $m=m_1 m_2$...
In order to construct the solution, define $M_i$ to be $m/m_i$ and observe that $M_i$ is clearly an integer.
Since $M_i$ and $m_i$ are co-prime, each $M_i$ has a multiplicative inverse $n_i$ satisfying $M_i n_i \equiv 1 \mod m_i$.
The solution to the linear system of congruences is then
\begin{align}
x = c_1 n_1 M_1+ c_2 n_2 M_2+...
\end{align}
As indicated in Algorithm \ref{alg:RowReduce}, \FFSol{} only ever uses the Chinese remainder theorem over two module $m_1$ and $m_2$.
In this case, $M_1 =m_2$ and $M_2=m_1$.
The Euclidean algorithm then provides $n_1$ and $n_2$ satisfying $m_2 n_1+m_1+n_2=1$.
The solution $x$ is simply given by
\begin{align}
x = c_1 n_1 m_2 + c_2 n_2 m_1.
\end{align}

\section{Benchmarks of fully dense equations}
\label{sec:appendixDense}

\begin{table}
\centering
\caption{\FFSol{} was benchmarked against Mathematica in solving a 100\% dense matrix.
The ratios in the table are with respect to the time and memory it takes \FFSol{} to solve the system.}
\subcaption*{Benchmarks on MacOS with i5-7267U, 16GB memory, Mathematica 13.2, and GCC 12.2}
\begin{tabular}{l |c|c|c|c}
	& Time (s) & Time ratio & Memory (GB) & Memory ratio\\
\hline
\FFSol{} & 63 & 1 & 0.63 & 1\\
Mathematica \code{Solve} & 630 & 10 & 5.7 & 9.0\\
\LinSysSol{} & 99 & 1.6 & 2.3 & 3.7\\
\spasmlink{} & 120 & 1.9 & 1.1 & 1.7 \\
\begin{tabular}{@{}l@{}} \FiniteFlow{} \\  ~~~~ \ffdense{} \end{tabular} & 98 & 1.6 & 1.2 & 1.9 \\
\begin{tabular}{@{}l@{}} \FiniteFlow{} \\  ~~~~ \ffsparse{} \end{tabular} & 250 & 4.0 & 2.5 & 4.0
\end{tabular}
\bigskip
\subcaption*{Benchmarks on Ubuntu under WSL2 with i5-6600, 32GB memory, Mathematica 13.1, and GCC 9.4}
\begin{tabular}{l |c|c|c|c}
	& Time (s) & Time ratio & Memory (GB) & Memory ratio\\
\hline
\FFSol{} & 41 & 1 & 0.64 & 1\\
Mathematica \code{Solve} & 520 & 13 & 6.3 & 9.8\\
\LinSysSol{} & 82 & 2.0 & 2.3 & 3.6\\
\kira{} \code{2} (\firefly{}) & 370 & 9.0 & 0.56 & 0.86 \\
\kira{} \code{2} (\fermat{}) & 91 & 2.2 & 0.32 & 0.50\\
\begin{tabular}{@{}l@{}} \FiniteFlow{} \\  ~~~~ \ffdense{} \end{tabular} & 73 & 1.8 & 1.3 & 2.0 \\
\begin{tabular}{@{}l@{}} \FiniteFlow{} \\  ~~~~ \ffsparse{} \end{tabular} & 190 & 4.6 & 2.5 & 3.9
\end{tabular}
\bigskip
\subcaption*{Matrix properties}
\begin{tabular}{cc}
Matrix dimensions & 2,000 $\times$ 2,001 \\
Matrix density & 1\\
Highest row density & 1
\end{tabular}
\label{tbl:benchmark3}
\end{table}

\begin{table}
\centering
\caption{
The table below gives the breakdown of where \FFSol{} spends most of its time when solving the system in Table \ref{tbl:benchmark3}.
The breakdown is virtually identical between machines.
Note that the total for the breakdown is not 100\% due to rounding and because \FFSol{} performs other small housekeeping tasks not mentioned in the table such as attempting to prune the matrix (see Section \ref{sec:technical}), testing the solution, etc.
}
Number of row reductions used to solve the system:  2

\vspace*{1em}

\begin{tabular}{l  l}
Time breakdown: & \\
\hline
Row reduction & 40\% \\
Extended Euclidean algorithm & $<$1\% \\
Chinese remainder theorem & $<$1\% \\
Converting equations to matrix & 50\%
\end{tabular}
\label{tbl:breakdown3}
\end{table}

The two examples presented in Section \ref{sec:example} involve only moderately dense systems of equations.
Since \FFSol{} is designed to be able to solve fully dense equations, it is informative to benchmark the solver on such a system.
The example benchmarked in this section is a 100\% dense system of 2,000 equations.
This is enough equations to demonstrate the advantages of \FFSol{} without being too computationally intensive.
The matrix is engineered so that it has exactly one null vector, the entries of which are random rational numbers with numerators and denominators less than $10^3$.
The entries of the unsolved matrix are also random rational numbers where many, but not all, of the numerators and denominators are less than $10^3$.
The detailed construction of this matrix can be found in the example file accompanying \FFSol{}.
This matrix was designed to be fully dense but has a relatively simple solution in order to mimic a real-world physics problem.
For example, imposing factorization on the ansatz for some amplitude by numerically sampling random kinematics can generate systems of equations with a similar character to the system constructed here.

The results of the benchmark are presented in Table \ref{tbl:benchmark3}.
\FFSol{} was timed against Mathematica's \code{Solve}, \LinSysSol{}, \spasmlink{}, \kira{}, and \FiniteFlow{}.
This example is now firmly outside of the intended use case of many of the sparse solvers.
In keeping with previous examples, \FFSol{} remains the fastest option.
\kira{} with \fermat{} is the only configuration that beats \FFSol{} in terms of memory, using roughly half as much memory at the cost of being approximately two times slower.
\FFSol{} uses about an order of magnitude less memory than \code{Solve}, just like the DBI example.
Unlike the DBI example where \FFSol{} was roughly two orders of magnitude faster than \code{Solve}, here it is only about one order of magnitude faster.
The reason for the speed discrepancy may be that \FFSol{} makes certain rudimentary optimizations for sparser systems like skipping rows with zero in the pivot column.
For a fully dense system these zeros will generally never occur, resulting in diminished performance.
As shown in Table \ref{tbl:breakdown3}, converting the equations into a matrix now takes the majority of the time because of the density of the system.
Row reduction still occupies a large fraction of the overall solve time because there are no rows with zero in the pivot column that can be skipped.
The solution to the system remains relatively simple so the extended Euclidean algorithm and the Chinese remainder theorem are computationally inexpensive compared to other steps.
Turning off OpenMP has virtually no impact on performance and failing to compile the prime only results in performance dropping by about 30\%.
Both of these observations might be explained by a memory bandwidth bottleneck.



\bibliographystyle{JHEP}
\bibliography{biblio.bib}







\end{document}